\documentclass{ws-procs9x6}
\usepackage{t1enc}
\usepackage[latin2]{inputenc}
\usepackage[english]{babel}
\usepackage{graphicx,color,epsfig,float,amsmath,hyperref,multirow}
\DeclareGraphicsExtensions{.pdf,.png,.jpg,.svg}
\graphicspath{{figures/}}

\begin{document}

\title{Time evolution of the anisotropies of the hydrodynamically expanding sQGP}

\author{A. BAGOLY, M. CSAN\'AD$^*$}

\address{Department of Atomic Physics, E\"otv\"os University,\\
Budapest, P\'azm\'any P. s. 1/a, H-1117, Hungary\\
$^*$E-mail: csanad@elte.hu}

\begin{abstract}
In high energy heavy ion collisions of RHIC and LHC, a strongly interacting quark gluon plasma (sQGP) is created. This
medium undergoes a hydrodynamic evolution, before it freezes out to form a hadronic matter. The initial state of the sQGP
is determined by the initial distribution of the participating nucleons and their interactions. Due to the finite number of 
nucleons, the initial distribution fluctuates on an event-by-event basis. The transverse plane anisotropy of the initial state
can be translated into a series of anisotropy coefficients or eccentricities: second, third, fourth-order anisotropy etc.
These anisotropies then evolve in time, and result in measurable momentum-space anisotropies, to be measured with
respect to their respective symmetry planes. In this paper we investigate the time evolution of the anisotropies.
With a numerical hydrodynamic code, we analyze how the speed of sound and viscosity influence this evolution.
\end{abstract}


\section{Hydrodynamics in ultra-relativistic heavy ion collisions}
The medium created in ultra-relativistic heavy ion collisions is successfully modeled by perfect fluid
hydrodynamics, i.e. the observables of soft hadrons, photons and leptons can described by hydro
models. Exact solutions are particularly interesting, since one gains an analytic insight on the connection between
the initial and final state of the hydro evolution.

These solutions fulfill a simple set of laws, governing many systems on many scales. Hydrodynamics
only assumes the local conservation of energy and momentum, and, depending on the circumstances, of
entropy or some conserved charges. The equations of non-relativistic hydrodynamics are:
\begin{align}
\frac{\partial \rho}{\partial t} + \bm{\nabla}\rho\bm{v}&=0\\
\rho\left(\frac{\partial \bm{v}}{\partial t}+(\bm{v\nabla})\bm{v}\right)&=-\bm{\nabla}p +\mu\Delta\bm{v}+\left(\zeta+\frac{\mu}{3}\right)\bm{\nabla}(\bm{\nabla v})\\
\frac{\partial \varepsilon}{\partial t}+\bm{\nabla}\varepsilon\bm{v}&=-p\bm{\nabla v}+\bm{\nabla}(\sigma {\bm v})
\end{align}
where $\rho$ is matter density, $\bm{v}$ is the velocity field, $\varepsilon$ is the (internal) energy density, $p$ is pressure, 
$\sigma$ is the viscous stress tensor of the homogeneous and isotropic medium, $\mu$ is the dynamic viscosity coefficient, $\lambda$ the bulk viscosity coefficient,
while $t$ represents time and $\bm{\nabla}=(\partial_x,\partial_y,\partial_z)$ the spatial derivative vector.
If we however would like to utilize hydrodynamics in a relativistically expanding medium, we have to use the equations
of relativistic hydrodynamics. In case of perfect hydro, these take the form of
\begin{align}
T^{\mu\nu}=\big(\varepsilon+p\big)\frac{u^\mu u^\nu}{c^2}-pg^{\mu\nu},\quad \partial_\mu T^{\mu\nu}=0.
\end{align}
To close this set of equations, we need to specify a relation between energy density $\varepsilon$ and pressure $p$. This
we obtain using the below Equation of State:
\begin{equation}
\varepsilon=\kappa(T)p,
\end{equation}
where $\kappa=c_s^{-2}$ is the inverse square of speed of sound. It's temperature dependence can be taken from lattice
QCD calculations, but often $\kappa$ is simply kept constant.

\section{Initial conditions and anisotropies}

\begin{figure}
\centering
	\includegraphics[width=0.4\textwidth]{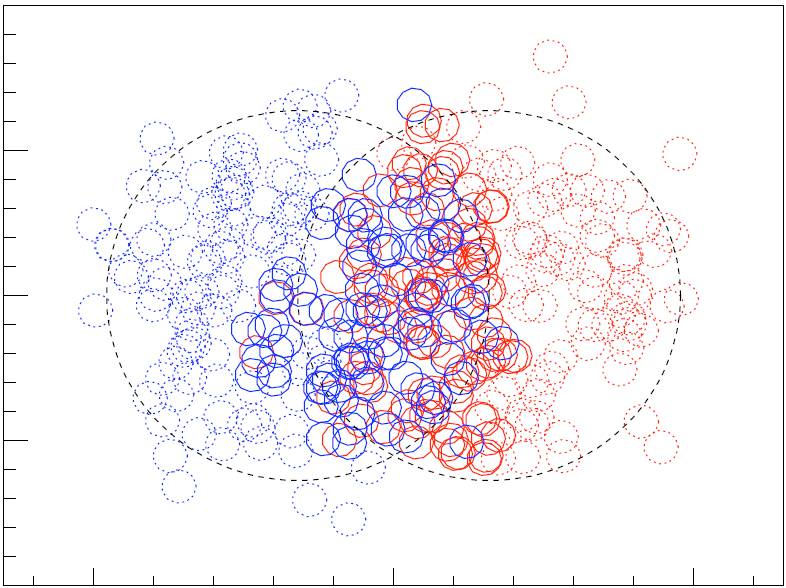}
	\caption{Simulated participant nucleon distribution (figure from Ref.~\cite{Loizides:2014vua})}
	\label{fig:ini_mc}
\end{figure}

There are many solutions that fulfill the above equations, we however need to start from an initial state that corresponds
to the energy distribution deposited by the nucleons participating in the collision.
In nucleus-nucleus collisions the initially formed medium can be spatially approximated by an ellipsoid, which can be
described by the contour levels of a scale variable $s$
\begin{align}
s=\frac{x^2}{X^2}+\frac{y^2}{Y^2}+\frac{z^2}{Z^2}
\end{align}
where $X,Y,Z$ are the axes of the ellipsoid described by the equation $s=1$. Multiple exact solutions of hydrodynamics were
found~\cite{Csorgo:2003ry,Csorgo:2001xm} that resemble this symmetry, i.e. where the thermodynamical quantities are
constant on $s=$const. curves. It is important to see however, that nuclei contain a finite number of nucleons, and thus
the participants of a nucleus-nucleus collision form an event-by-event fluctuating shape (see Fig.~\ref{fig:ini_mc}).

\begin{figure}
\centering
	\includegraphics[width=0.8\textwidth]{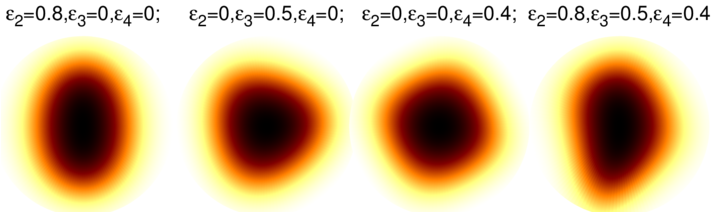}
	\caption{Initial distribution with the outlined $s$ scaling variable, for different $\varepsilon_n$ values.}
	\label{fig:ini_hydro}
\end{figure}

Thus a simple ellipsoidal symmetry is not sufficient to describe the hydrodynamically expanding medium, not to mention
that it does not describe higher order flow coefficients either. To overcome this difficulty, one can assume higher order 
anisotropies in the scaling variable, by redefining it as
\begin{align}
s=\frac{r^2}{R^2}\left(1+\varepsilon_2\cos(2\phi)\right)+\frac{z^2}{Z^2}
\end{align}
with $r$ and $\phi$ being the polar coordinates in the transverse plane, while $R$ and $\varepsilon_2$ are in a direct
relation to $X$ and $Y$. This can then be generalized to
\begin{align}
s=\frac{r^2}{R^2}\left(1+\sum\limits_{n=2,3,\dots}\varepsilon_n\cos{n\varphi}\right)+\frac{z^2}{Z^2}
\end{align}
in the $\cos{n\varphi}$ expression a $\Psi_n$ $n$-th order event plane can also be introduced. Example 
distributions generated with this scaling variable are shown in Fig.~\ref{fig:ini_hydro}. Solutions
utilizing scaling variables like the above $s$ were presented in Ref.~\cite{Csanad:2014dpa}. However, these
solutions, as well as the ones in Ref.~\cite{Csorgo:2003ry} are valid only in case of a 3D Hubble-flow, which
corresponds to the lack of pressure gradients. This also means that the anisotropies will remain constant 
in time, which may not be the case throughout the whole evolution. In our analysis, we thus apply numerical calculations
to include pressure gradients in the initial conditions, and analyze the time evolution of the anisotropies.

Thus we will start the hydrodynamic evolution from the above described set of initial conditions. We would like
to extract the anisotropy coefficients at later stages of the evolution, for which we need a definition.
We introduced the anisotropies of density, energy density and velocity fields as follows: 
\begin{align}
\epsilon_n=\langle \cos(n\varphi)\rangle_{\rm \rho/\varepsilon/w},\label{eq:a1}
\end{align}
where $w=\exp{(-v_x^2-v_y^2)}$ is defined to calculate the asymmetry of velocity field.
To understand the relation of the asymmetry parameters introduced by \ref{eq:a1} to the scale variable anisotripy
coefficients, $\varepsilon_n$'s, we checked their relation.  As a first order approximation (in $\varepsilon_n$), we obtained
\begin{align}
\epsilon_n = -\varepsilon_n/2
\end{align}
however, in a more accurate approximation the result is:
\begin{align}
\epsilon_1&\approx \frac{(\varepsilon_2+\varepsilon_4)\varepsilon_3}{2+\sum_{n}\varepsilon_n^2}
=\frac{\epsilon_3(\epsilon_2+\epsilon_4)}{2}+\mathcal{O} (\epsilon_n^4)\\\label{eq:epsvareps1}
\epsilon_2&\approx\frac{-\varepsilon_2+\varepsilon_2\varepsilon_4}{2+\sum_{n}\varepsilon_n^2}
=-\frac{\epsilon_2}{2} +  \frac{\epsilon_2\epsilon_4}{2}+
\frac{1}{4}\epsilon_2\sum_n\epsilon_n^2 +\mathcal{O}  (  \epsilon_n^4 )\\
\epsilon_3&\approx\frac{-\varepsilon_3}{2+\sum_{n}\varepsilon_n^2}
=-\frac{\epsilon_3}{2} +  
\frac{1}{4}\epsilon_3\sum_n\epsilon_n^2 +\mathcal{O}  (  \epsilon_n^4 )\\
\epsilon_4&\approx\frac{-\varepsilon_4+\frac{1}{2}\varepsilon_2^2}{2+\sum_{n}\varepsilon_n^2}
=-\frac{\epsilon_4}{2} +   \frac{\epsilon_2^2}{4}-
\frac{1}{4}\epsilon_4\sum_n\epsilon_n^2 +\mathcal{O}  (  \epsilon_n^4 )\label{eq:epsvareps4}
\end{align}
this means that even if there is no $\varepsilon_1$ coefficient, we will still get an $\epsilon_1$. It is also
important to note that $\varepsilon_2\neq 0$ and $\varepsilon_{n\neq 2}=0$ will result in a
nonzero $\epsilon_4=\varepsilon_2^2/4$. See an example with multiple nonzero $\varepsilon_n$ coefficients in
table~\ref{t:eps}, where we give an example how a set of $\varepsilon_n$ values result in a different set
of $\epsilon_n$ values. The actual results can be accurately approximated with
eqs.~(\ref{eq:epsvareps1})-(\ref{eq:epsvareps4}).

\begin{table}
\tbl{Example for the $\epsilon_n -\varepsilon_n$ relation.}
{\begin{tabular}{@{}ccccc@{}}\toprule
\multirow{2}{*}{$n$} &
\multirow{2}{*}{$\varepsilon_n$} &
\multicolumn{3}{c}{$\epsilon_n$}\\
      &                         &  direct calculation & approximation & first order estimate\\
\colrule
1 & 0 & 0.00278 & 0.00298 & 0 \\
2 & 0.1 & -0.04927 & -0.04869 & -0.05 \\
3 & 0.05 & -0.02533 & -0.02484 & -0.025 \\
4 & 0.02 & -0.00769 & -0.00745 & -0.01 \\
\botrule
\end{tabular}}\label{t:eps}
\end{table}

\section{Numerical method}
Let us now discuss how we evolve the previously defined initial conditions in time. In order to do that, we
first transform the equations of (1+2 dimensional) hydrodynamics to a convective form as usual:
\begin{equation}
\partial_t {\bf Q}+\partial_x {\bf F}({\bf Q})+\partial_y {\bf G}({\bf Q})=0\label{eq:e1}
\end{equation}
where $\bf Q$ is calculated from the hydrodynamical fields, while $\bf F$ and $\bf G$ are functions of
these, transformed from the original hydro equations. The finite volume method discussed below and introduced
in Ref.~\cite{Toro:2006aa} works for many different systems of partial differential equations (PDE's), one just
needs to determine the given functions to transform the PDE's to the above form. This can be done in a straightforward
manner, see for example Refs.~\cite{Toro:2006aa,Nishikawa:2008aa,Takamoto:2011wi,Karpenko:2013wva}.

The basis of finite volume methods is to discretize the ${\bf Q}$ vector, by defining it on a space-time grid,
and obtaining the discrete values by averaging in each cell. Then, we
have to evaluate fluxes $\bf F$ and $\bf G$ between grid points (these are then called intercell fluxes). This is difficult, since the
fluxes depend on the values of $\bf Q$, but these are defined on the grid points. Our method, outlined in
Ref.~\cite{Toro:2006aa} and used for exampe in Ref~\cite{Takamoto:2011wi}, is based on an accurate multi-stage (MUSTA)
predictor-corrector estimation of the intercell fluxes. We discuss this method briefly in 1+1 dimensions (and with a
single-component $Q$ only) -- it is straightforward to generalize it to multiple spatial dimensions via operator splitting (we
used the Lie type of splitting method here). In case of viscous hydrodynamics, we also used operator splitting to separate
ideal and viscous fluxes.

To estimate the intercell fluxes between cells $i$ and $i+1$, we start from the $Q^n_i$ and $Q^n_{i+1}$ values,
where $n$ represents the current time-step. Let us call these values $Q^{(0)}_{i,i+1}$, indicating that these
represent a ``zeroth order'' estimate of the intercell values. We will then make a series of estimates in the given cell,
the $l$th estimate being indicated with $Q^{(l)}_{i,i+1}$. The flux in the cell centers is then
$F^{(l)}_{i,i+1} \equiv F\left(Q^{(l)}_{i,i+1}\right)$. We define an intermediate $Q$ value and flux $F$ as follows:
\begin{align}
Q^{(l)}_{i+\frac{1}{2}}=\frac{1}{2}\left[Q^{(l)}_i+Q^{(l)}_{i+1}\right]-\frac{1}{2}\frac{\Delta t}{\Delta x}\left[F^{(l)}_{i+1}-F^{(l)}_i\right], \quad F_M^{(l)} \equiv F\left(Q^{(l)}_{i+\frac{1}{2}}\right)
\end{align}
Then our intercell flux estimate is
\begin{align}
F^{(l)}_{i+\frac{1}{2}}=\frac{1}{4}\left[F^{(l)}_{i+1}+2F^{(l)}_M+F^{(l)}_{i}-\frac{\Delta x}{\Delta t}\left(Q^{(l)}_{i+1}-Q^{(l)}_i\right)\right]
\end{align}
This is the so called FORCE flux approximation. The essence of our the applied MUSTA algorithm is that now we make a new
prediction, $Q^{(l+1)}_i$ and with that we repeat the steps and we get a better flux approximation, and with that we can make
better prediction, and again better flux approximation. To make a new prediction simply we use the equations we want to solve:
\begin{align}
Q_i^{(l+1)}&=Q^{(l)}_i-\frac{\Delta t}{\Delta x}\left[F^{(l)}_{i+\frac{1}{2}}-F^{(l)}_i\right]\\
Q_{i+1}^{(l+1)}&=Q^{(l)}_{i+1}-\frac{\Delta t}{\Delta x}\left[F^{(l)}_{i+1}-F^{(l)}_{i+\frac{1}{2}}\right]
\end{align}
We may then stop at a given $l$ value, by setting $F^{n}_{i+\frac{1}{2}} = F^{(l)}_{i+\frac{1}{2}}$. With this, we simply
calculate the next time-step as
\begin{align}
Q_i^{n+1} = Q_i^n - \frac{\Delta x}{\Delta t}\left(F^{n}_{i+\frac{1}{2}} - F^{n}_{i-\frac{1}{2}} \right).
\end{align}
where $F^{n}_{i-\frac{1}{2}}$ can also be calculated by the procedure outlined above, if we start from cells $i-1$ and $i$.

In order to complete the method, let us mention that we used the initial conditions as described above,
we used non-reflective boundary conditions, and the timestep was determined adaptively via the
Courant-Friedrichs-Lewy condition. We tested our method with the usual Sod shock tube and known hydrodynamic
solutions as well.

\section{Results}
With the above outlined method, we solved the equations of nonrelativistic hydrodynamics with viscosity, as well as
the equations of relativistic perfect hydrodynamics. Our goal was to investigate how various speed of sound values
and various viscosity values affect the time evolution of the asymmetries.

\begin{figure}
	\centering
    		\includegraphics[width=0.49\textwidth]{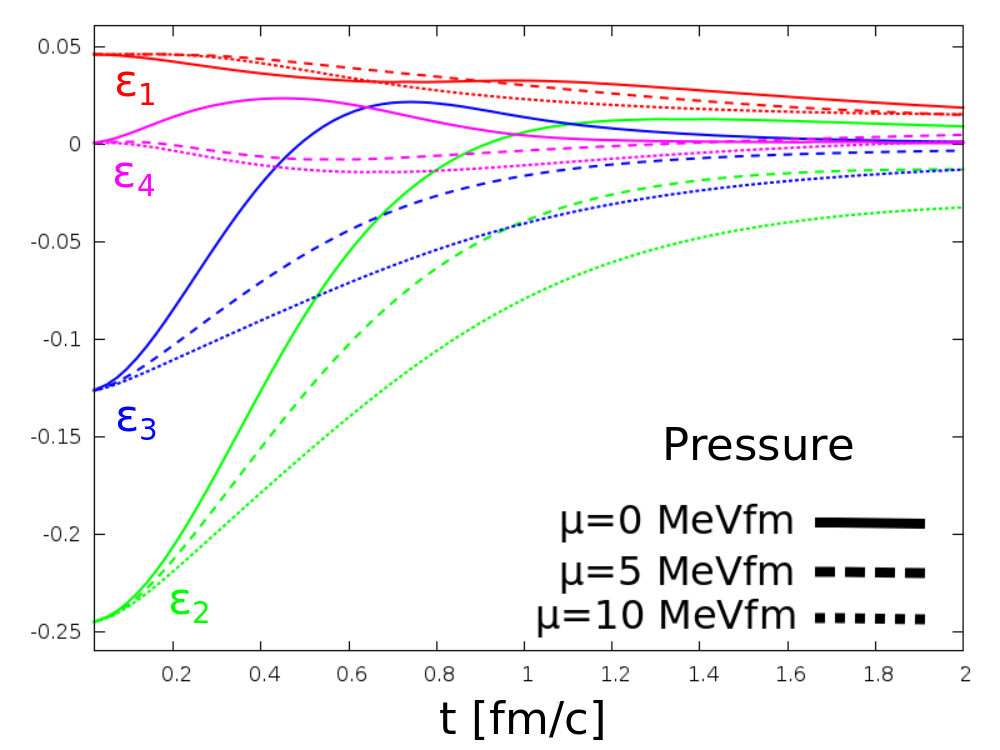}
        	\includegraphics[width=0.49\textwidth]{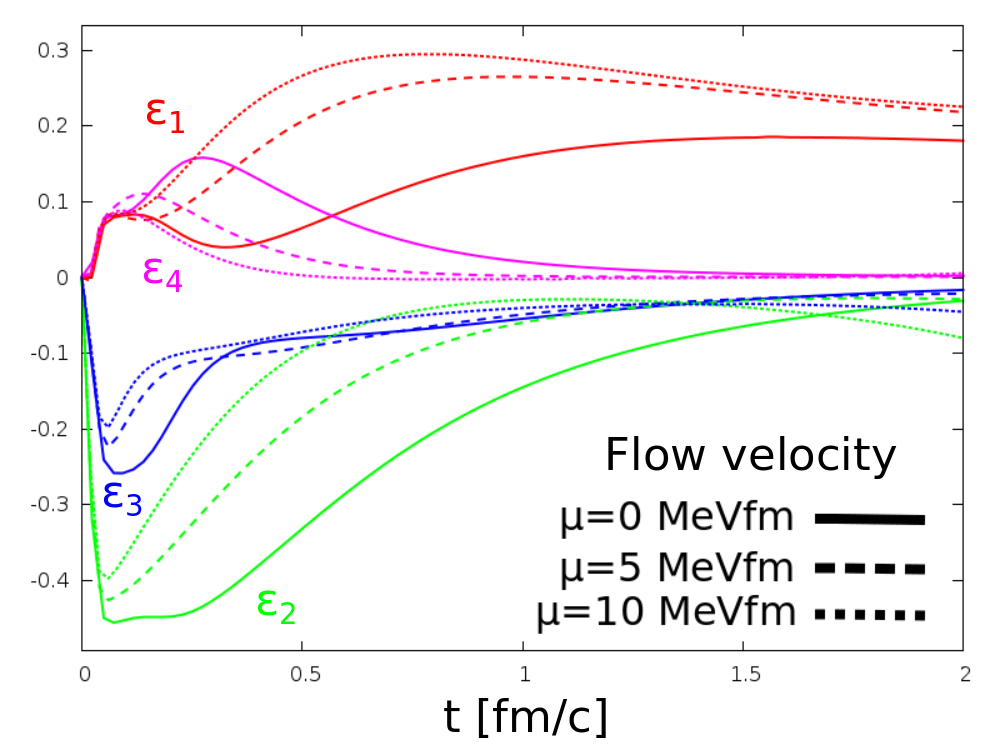}
	\caption{Time evolution of pressure anisotropies is shown in the left panel, while anisotropies in
	the flow field are shown in the right panel.}
	\label{fig:nonrel_eps_visc}
\end{figure}

\begin{figure}
	\centering
    		\includegraphics[width=0.99\textwidth]{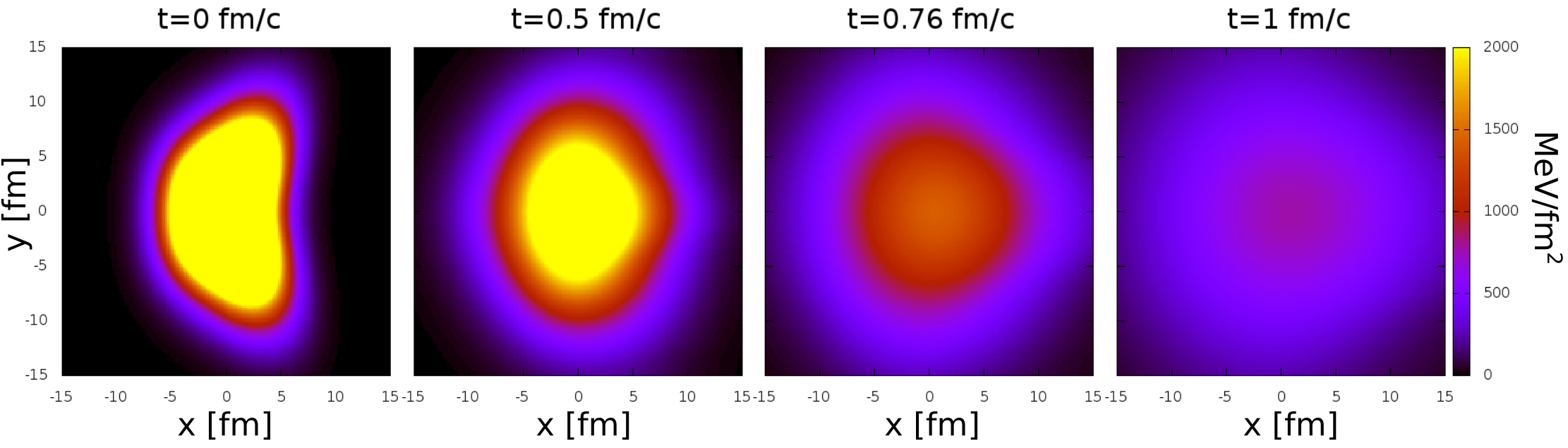}
    		\includegraphics[width=0.99\textwidth]{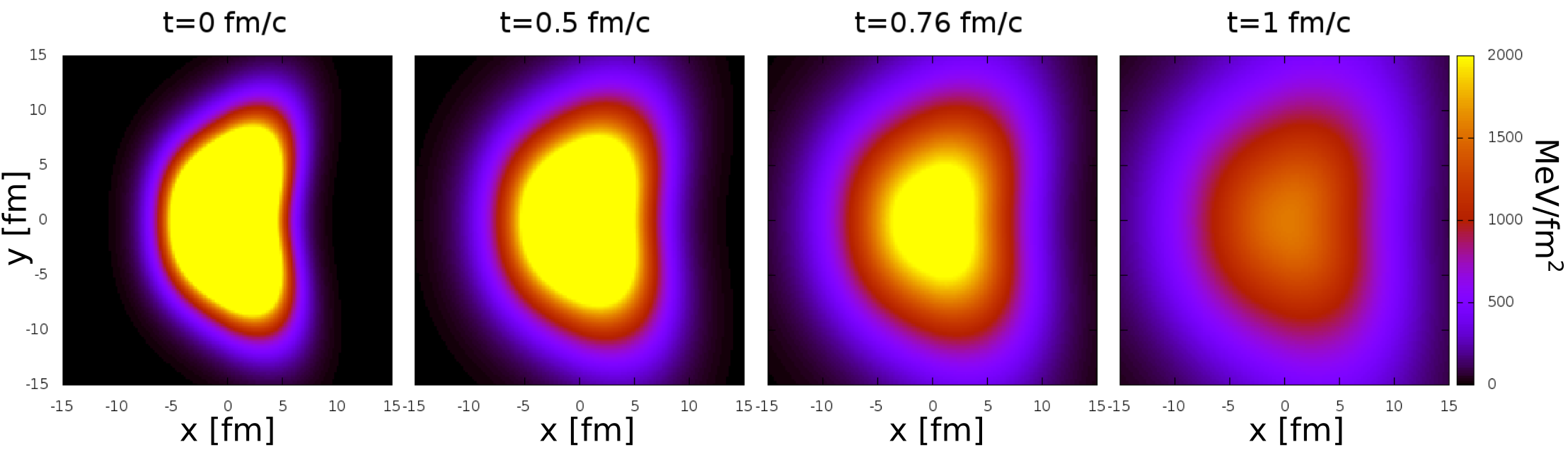}
    		\includegraphics[width=0.99\textwidth]{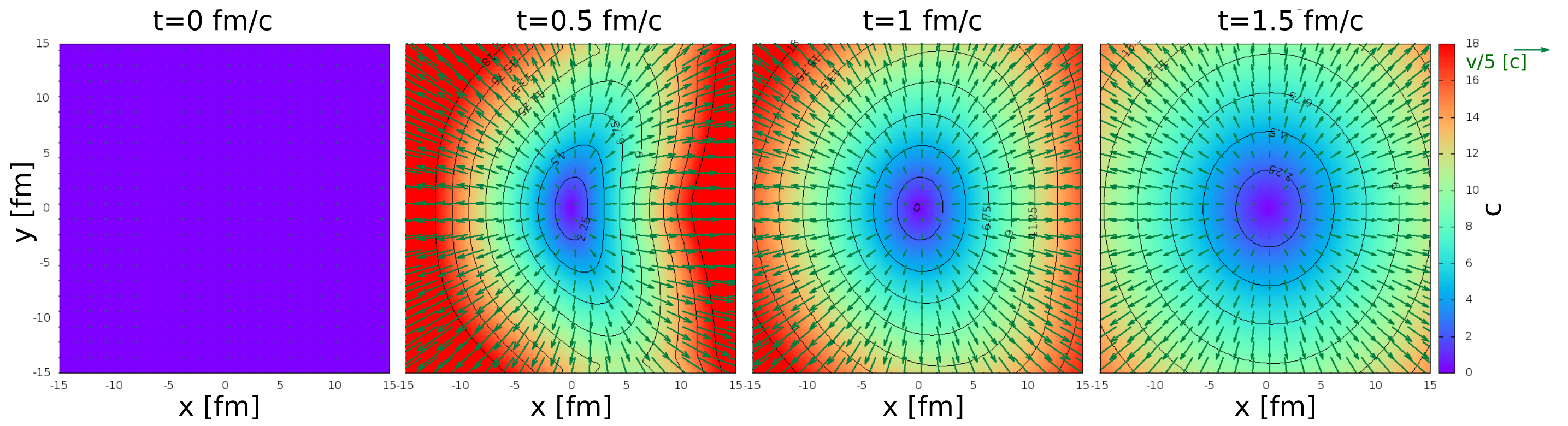}
    		\includegraphics[width=0.99\textwidth]{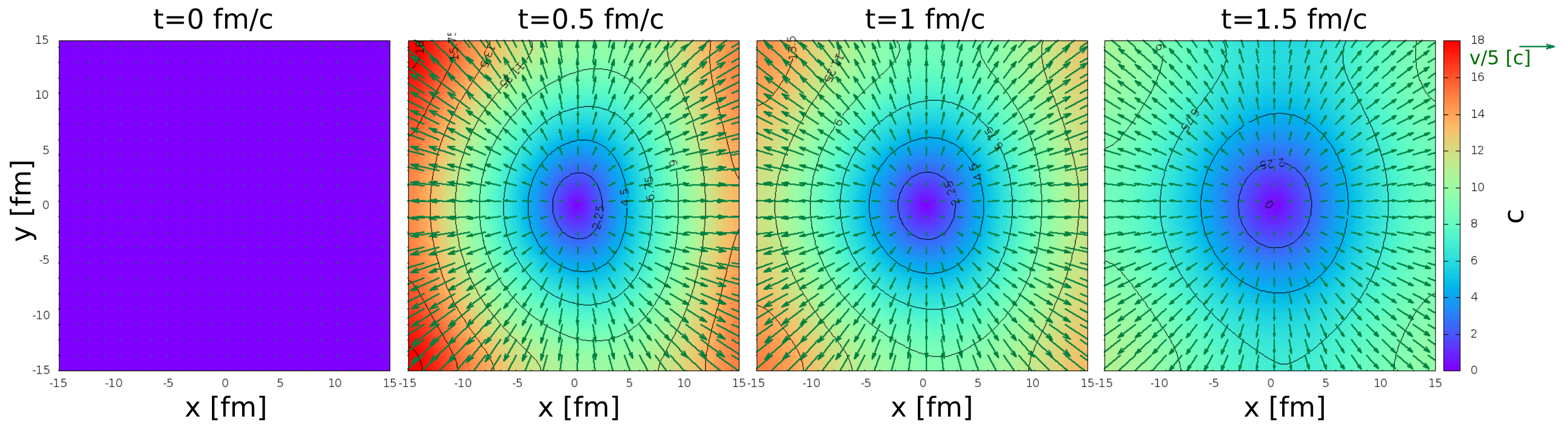}
	\caption{Time evolution of the energy density (first two rows) and of the flow field (last two rows). The first and third
	rows show the viscosity free case, the second and fourth rows show the viscous case with
	$\mu={\rm 10MeV}\cdot{\rm fm}$.}
	\label{fig:nonrel_anim_visc}
\end{figure}

Let us first discuss the effects of viscosity (in nonrelativistic hydro). As shown in Fig.~\ref{fig:nonrel_eps_visc}
viscosity makes the disappearance of pressure anisotropies slower, and the same is in the number density.
In other words, in case of nonzero viscosity the pressure asymmetries
remain in the system for a longer time. This may be due to the fact that viscosity makes the flow slower, as
illustrated in the first two plots of Fig.~\ref{fig:nonrel_anim_visc}. As shown in the bottom plot
of Fig.~\ref{fig:nonrel_eps_visc} and the third and fourth plots in Fig.~\ref{fig:nonrel_anim_visc}, the effect of
viscosity on the flow field is however the opposite: viscosity makes the flow anisotropies disappear faster.
The reason of this may be that the viscous force contains the second spatial derivatives, thus the fluid elements
with larger asymmetries feel a stronger viscous force, which means that viscosity ``washes out'' the anisotropies.

\begin{figure}
	\centering
    		\includegraphics[width=0.49\textwidth]{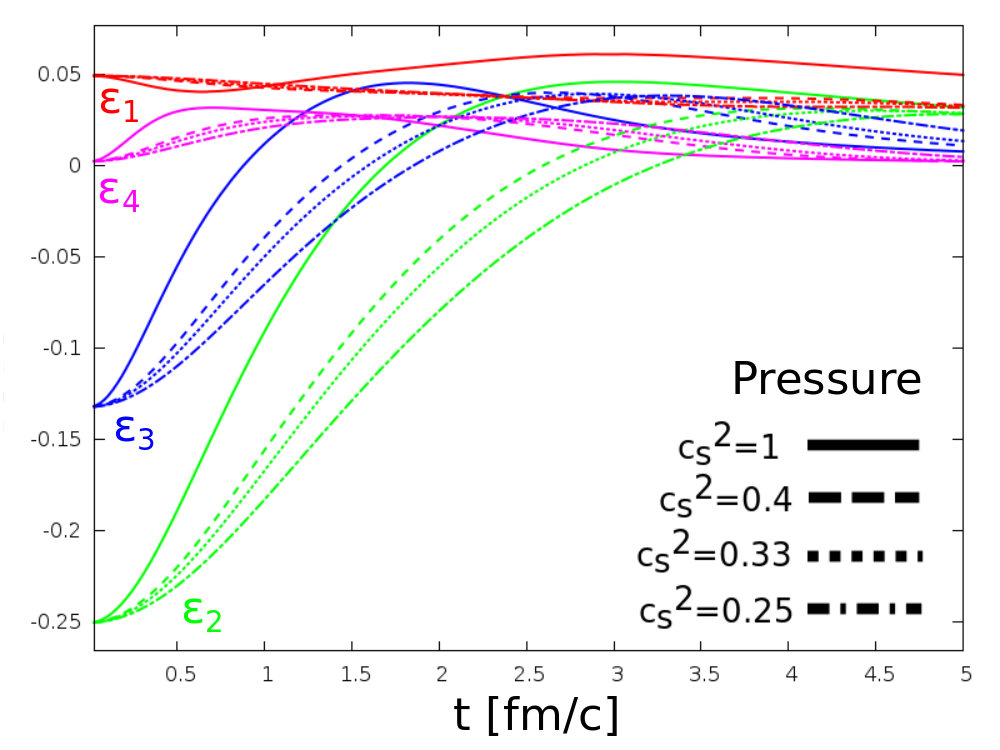}
        	\includegraphics[width=0.49\textwidth]{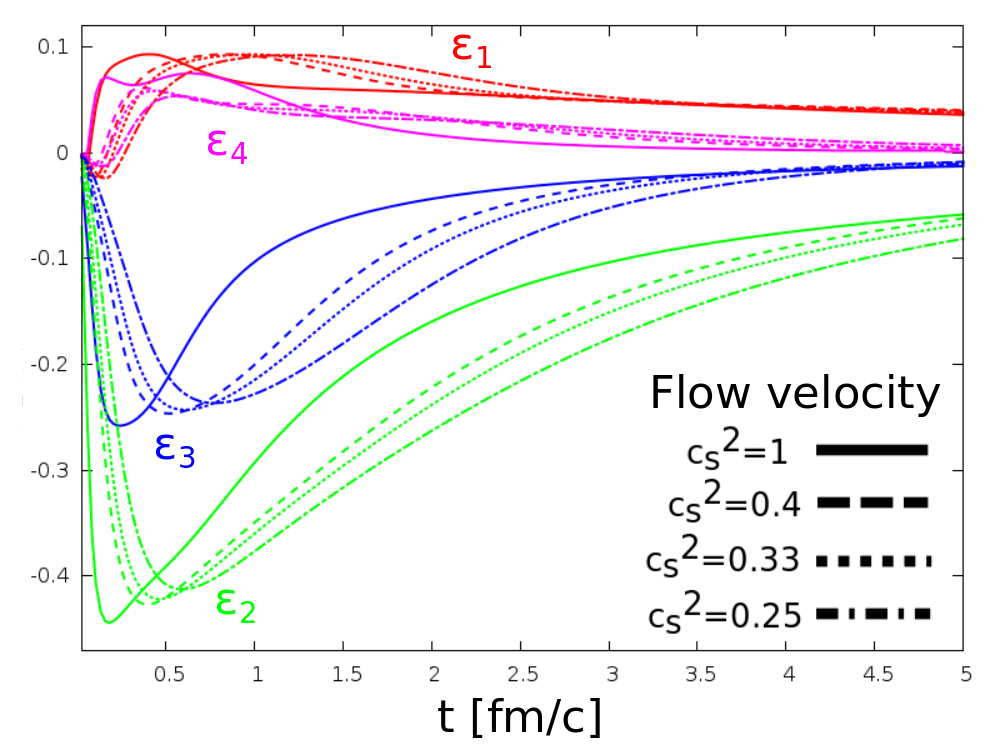}
	\caption{Time evolution of pressure anisotropies is shown in the left panel, while anisotropies in
	the flow field are shown in the right panel.}
	\label{fig:nonrel_eps_cs}
\end{figure}

\begin{figure}
	\centering
    		\includegraphics[width=0.49\textwidth]{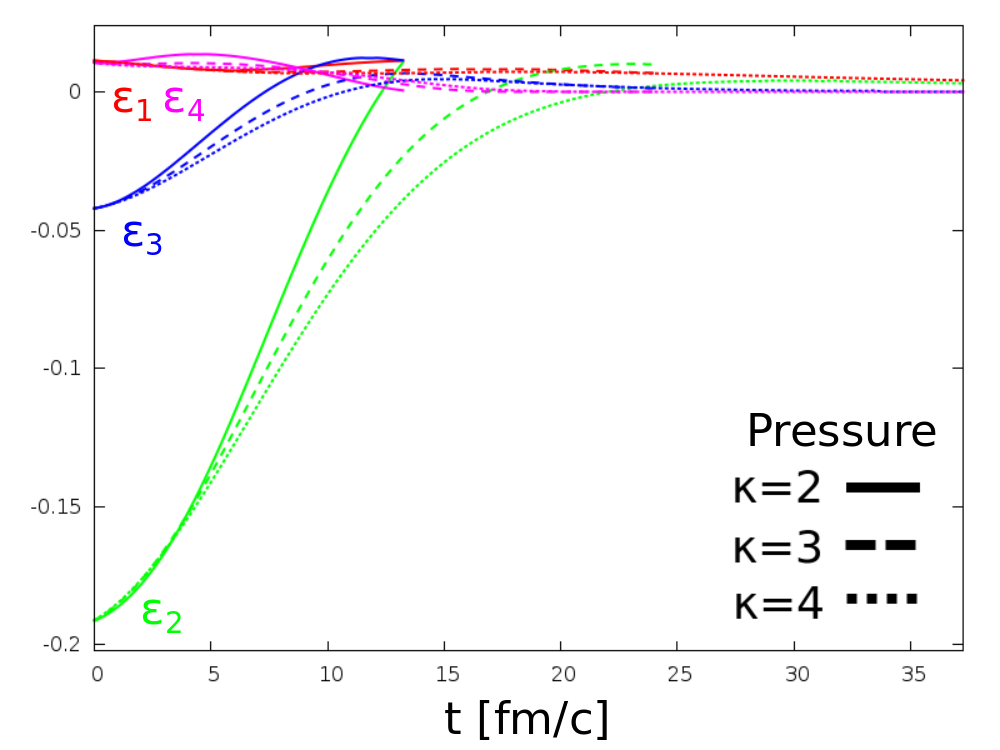}
    		\includegraphics[width=0.49\textwidth]{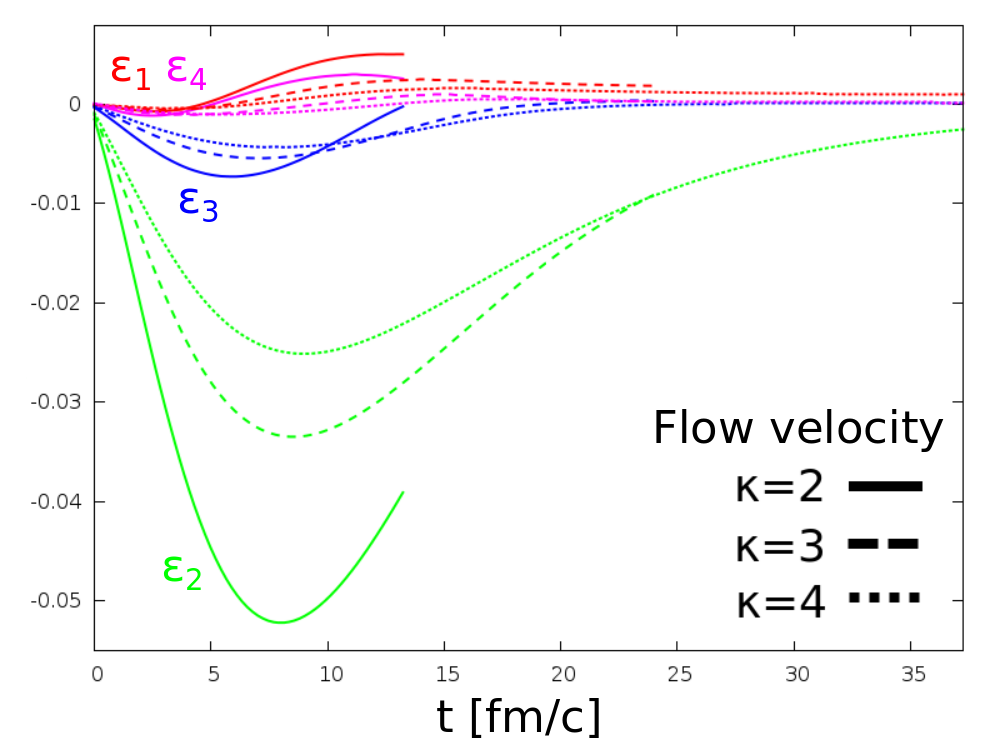}
    		\includegraphics[width=0.99\textwidth]{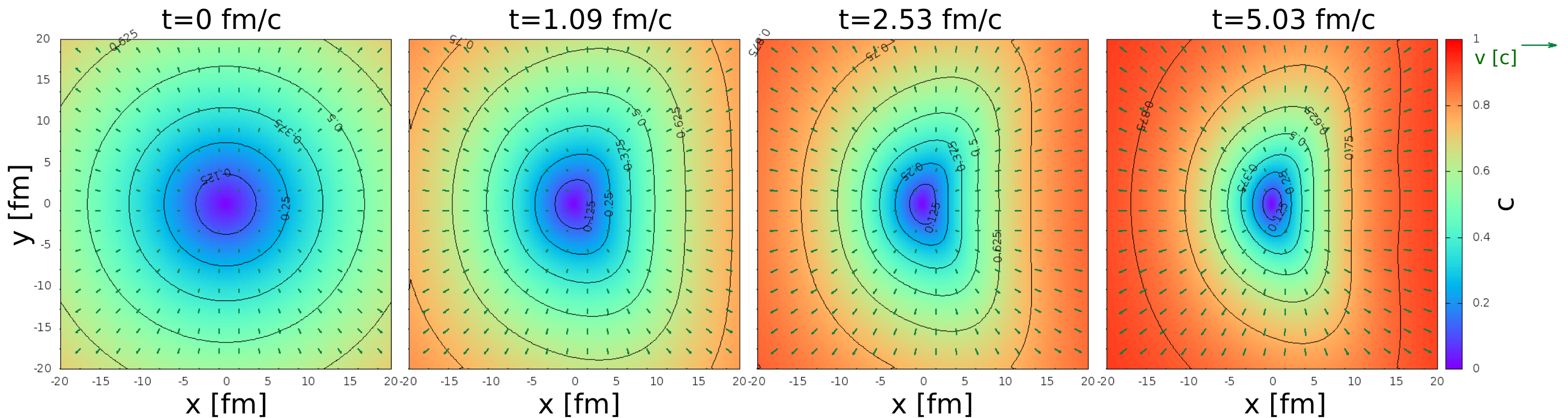}
    		\includegraphics[width=0.99\textwidth]{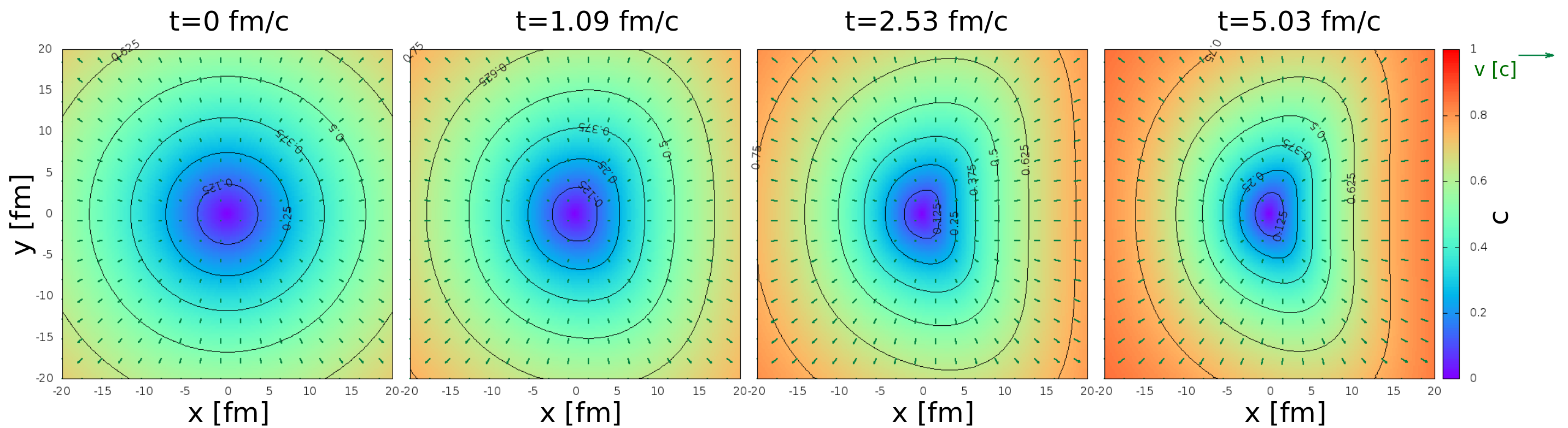}
	\caption{The time evolution of velocity field. The top row shows $\kappa=2$ case, the bottom row the $\kappa=4$ case.
	Time evolution of pressure anisotropies is shown in the left panel, while anisotropies in
	the flow field are shown in the right panel.}
	\label{fig:rel_eps_kappa}
\end{figure}

We also investigated the effects of a change in the speed of sound. As shown in Fig.~\ref{fig:nonrel_eps_cs},
the reduction of speed of sound makes the anisotropies disappear slower. The reason for this is that pressure
waves travel with the speed of sound, so the equalization of pressure is slower in case of smaller speed of sound
values. This is confirmed by relativistic calculations as well. As shown in the top panels of Fig.~\ref{fig:rel_eps_kappa},
in case of a larger $\kappa$ (i.e. a smaller $c_s^2$ value) the anisotropies disappear slower, and also
the cooling is slower. This also means that of course the system will freeze out later. The time evolution of the flow
field is shown in the bottom panels of Fig.~\ref{fig:rel_eps_kappa}.

\section{Summary}
In this paper we analyzed the time evolution of the asymmetries of the QCD matter created in high energy heavy ion
collisions. We utilized numerical hydrodynamics instead of exact solutions to investigate situations not describable
by the presently known hydrodynamic solutions. We however did not start from the most realistic initial conditions,
but started instead from an initial condition that is very similar to one described by known analytic solutions -- except
in pressure, where we used a pressure profile similar to the density profile given in these models. We analyzed both
non-relativistic and relativistic hydrodynamics, and arrived at similar conclusions. It turns out that a stiffer equation of
state, i.e. a larger speed of sound makes the pressure anisotropies disappear faster. The appearance of viscosity also
makes the flow anisotropies disappear faster, however, pressure anisotropies remain in the system longer, due to 
the slower flow of the viscous system.

\section*{Acknowledgments}
The authors would like to thank Julia Ny\'iri for her kind invitation to this series, and the organizers for keeping
Vladimir Gribov's work and memory alive. M. Cs. would also like to thank Tam\'as Cs\"org\H{o} for inspiring and
useful discussions. This manuscript was supported by the OTKA grant NK 101438.

\bibliographystyle{ws-procs9x6}
\bibliography{../../Master}

\end{document}